\documentclass{article}

\usepackage{amssymb}
\usepackage{graphicx}
\usepackage{amsmath}

\begin{document}

\author{Ad\'{e}lcio C. Oliveira \\ Departamento de Ci\^{e}ncias Exatas e
Tecnol\'{o}gicas\\ Universidade Estadual de Santa Cruz, Ilh\'{e}us,
45662-000, BA, Brazil }

\title{  Short Time Dynamics of Scalar Products
in Hilbert Space}
\maketitle
%\pacs{03.65.Ca,03.65.Sq,03.65.Ta,03.65.Ud,03.65.Wj}
%\keywords{Ehrenfest time, classical limit,Overlap, Fidelity Decay, Scars,
%Husimi Function, Chaos }

\begin{abstract}
\bigskip We use the semiclassical method proposed in \cite{Adelcio2003} to
study scalar products such as the overlap, Husimi functions and
fidelity decay. Scars of classical periodic orbits arise naturally
in this pertubative expansion. We also derive analytically a well
known numerical result that fidelity has a quadratic decay for short
times. We study the overlap in the chaotic regime and integrable for
some simple systems.
\end{abstract}

%\newpage

\section{Introduction}

Since early times of quantum theory, some quantization difficulties
 of non integrable systems were pointed by Einstein
\cite{Einstein,Aguiar}. Recently, due to the pioneer discoveries of
classically chaotic systems, the subject has yielded many
interesting and important results both from the point of view of
numerical models and (not as many) analytical proofs
\cite{Gutz,Haake,Stockmann}.

One interesting discovery initiated by Bogomolny\cite{Bogomoln86,
Bogomolny1988} and Heller\cite{Heller84, HellerBook} drew much
attention. They showed that the Hamiltonians eigenfunctions of
chaotic systems exhibit \textquotedblleft scars\textquotedblright\
around the periodic orbit. The effect is more conspicuous for
integrable systems than for chaotic ones, given the wealth of periodic orbits in the latter \cite%
{Oliveira01}. The search for classical \textquotedblleft
imprints\textquotedblright\ than the celebrated phenomena of scars
on eigenfunctions of quantum systems with classical analog has also
gained a lot of attention. It was suggested \cite{Bohigas} that a
 quantum spectra density and classical behavior are correlated .
 It was been demonstrated the existence of long-range correlation in quantum spectra and the existence of
periodic orbits in the classical chaotic system \cite{Wintgen}.
Recently a relation between quantum phase transition and a classical
instability points \cite{Heiss,Reis01,Reis02} was shown. All these
cited works have a common characteristic, they show the existence of
a close relationship between classical and quantum signatures.

We begin by  generalizing  the semiclassical expansion
\cite{Adelcio2003} for n-dimensional system. The semiclassical
expansion is builded such as the first order wave function contains
the classical dynamics for the system in question as completely as
possible, in the sense that the dominant term is given only in terms
of classical trajectories. All next to first order contributions
contain essentially quantum effects and allow for a precise
identification of departure from classical behavior in the quantum
dynamics for short times. Secondly we use this expansion to extract
classical ingredients known to be contained in several scalar
products of states in the Hilbert space. We show in a very simple
way how the phenomena of scars naturally arises from an adequate
semiclassical analysis. On the other hand Husimi distributions and
fidelity can be treated along the same lines and well known results
from numerical studies can be given an analytical basis. The main
advantage of the proposed method is due to his simplicity and his
long domain of application, contrasting with the most actually used
methods\cite{Gutz, Haake,Stockmann}.

\section{\protect\bigskip The Semiclassical Expansion}

 Let us consider a classical one degree of freedom Hamiltonian of the form
\begin{equation}
H_{\text{cls}}=\frac{p^{2}}{2m}+V(q), \label{2.1}
\end{equation}
where $p$ stands for the particle momentum and $q$ for its position.
We make a change of variables
\begin{equation}
p=\frac{\alpha -\alpha ^{\ast }}{i\sqrt{\frac{2}{m\omega \hbar }}};\qquad q=%
\frac{\alpha +\alpha ^{\ast }}{\sqrt{\frac{2m\omega }{\hbar }}},
\label{2.2}
\end{equation}
where $\omega =\sqrt{k/m}$ and $k=\frac{\partial ^{2}V(q)}{\partial q^{2}}%
\rvert _{q=0}$.

The Hamiltonian can then be rewritten as
\begin{equation} H_{\text{cls}}=\hbar \omega \,\alpha ^{\ast }\alpha +U(\alpha ^{\ast }\alpha
)
\end{equation}
with $U(\alpha ^{\ast }\alpha )=V(q)-[k(\alpha +\alpha ^{\ast })/\sqrt{%
2m\omega /\hbar }]^{2}$.

We can write  $H_{\text{cls}}$ as a Taylor expansion, as
\begin{equation*}
H_{\text{cls}}=\hbar \omega \,\alpha ^{\ast }\alpha
+\sum_{m,n}A_{m,n}\left( \alpha ^{\ast }\right) ^{m}\alpha ^{n},
\end{equation*}
where $A_{1,1}=0$.

The classical equations of motions read
\begin{align}
\frac{d}{dt}\alpha & =\frac{1}{i\hbar }\frac{\partial H_{\text{cls}}}{%
\partial \alpha ^{\ast }}=-i\omega \alpha -\frac{i}{\hbar }%
\sum_{m,n}m\,A_{m,n}\left( \alpha ^{\ast }\right) ^{m-1}\alpha ^{n},
\label{2.3a} \\
\frac{d}{dt}\alpha ^{\ast }& =-\frac{1}{i\hbar }\frac{\partial H_{\text{cls}}%
}{\partial \alpha }=i\omega \alpha ^{\ast }+\frac{i}{\hbar }%
\sum_{m,n}n\,A_{m,n}\left( \alpha ^{\ast }\right) ^{m}\alpha ^{n-1}.
\label{2.3}
\end{align}
We choose the quantum Hamiltonian in order to have $\left\langle
\alpha \right| \hat{H}_{q}\left| \alpha \right\rangle
=H_{\text{cls}}$, if $\left| \alpha \right\rangle $ is a coherent
field state.

We make our semiclassical expansion around a quantum operator $%
H_{sc}(\alpha (t))$. The difference, $H-H_{sc}(\alpha (t))$, will be
considered as a perturbation .We choose the semiclassical
hamiltonian,  $H_{sc}(\alpha (t))$,  that for a coherent initial
state, all expectation values of point classical observables will be
precisely reproduced.

The semiclassical Hamiltonian which satisfies this condition
is\cite{Adelcio2003}:

\begin{equation}
\begin{split}
\hat{H}_{\text{sc}}(\alpha (t))& =\hbar \omega \,\hat{a}^{\dagger }\hat{a}%
+\sum_{m\neq 0}m\,A_{m,m}\left( \alpha ^{\ast }\right) ^{m-1}\alpha ^{m-1}(%
\hat{a}^{\dagger }\hat{a}-\alpha ^{\ast }\alpha
)+\sum_{m,n}A_{m,n}\left(
\alpha ^{\ast }\right) ^{m}\alpha ^{n} \\
& \qquad +\sum_{m\neq n}m\,A_{m,n}\left( \alpha ^{\ast }\right)
^{m-1}\alpha ^{n}(\hat{a}^{\dagger }-\alpha ^{\ast
})+\sum_{m,n}n\,A_{m,n}\left( \alpha ^{\ast }\right) ^{m}\alpha
^{n-1}(\hat{a}-\alpha ).
\end{split}
\label{9}
\end{equation}

\bigskip

In the case of several degrees of freedom the semiclassical
approximation is generalized as follows

\begin{eqnarray*}
\widehat{H}_{sc}(\alpha (t)) &=&H_{cl}+\widehat{H}_{sc}^{(1)}(\alpha
(t))\otimes \widehat{1}_{2}\cdot \cdot \cdot \otimes \widehat{1}_{N} \\
&&+\widehat{1}\otimes \widehat{H}_{sc}^{(2)}(\alpha (t))\otimes \widehat{1}%
_{3}\cdot \cdot \cdot \otimes \widehat{1}_{3}+\cdot \cdot \cdot
\end{eqnarray*}

\bigskip Consider one degree of freedom we can immediately write the semiclassical evolution
operator, just observing that
\begin{equation}
\hat{U}_{sc}(t)\left| \alpha _{0}\right\rangle =e^{i\phi (t)}\left| \alpha
(t)\right\rangle \Longrightarrow \hat{U}_{sc}(\alpha (t))=e^{i\phi
(t)}e^{i\Omega (t)a^{\dagger }a}\widehat{D}(\alpha (t))\widehat{D}(\alpha
_{0})^{-1}
\end{equation}
where $\widehat{D}$ stand for the well known displacement operator\ \

\begin{equation*}
\widehat{D}(\alpha )=e^{\alpha a^{\dagger }-\alpha ^{\ast }a}.
\end{equation*}

Thus, for the N dimensional case we have

\begin{eqnarray}
\hat{U}_{sc}(\alpha (t)) &=&\prod_{j}^{N}e^{i\phi (t)_{j}}e^{\alpha
_{j}(t)a_{j}^{\dagger }-\alpha _{j}^{\ast }(t)a_{j}}e^{-\alpha
_{j}a_{j}^{\dagger }+\alpha _{j}^{\ast }a_{j}}.  \label{USC} \\
&=&\prod_{j}^{N}\hat{U}_{sc}^{j}(\alpha (t)). \notag
\end{eqnarray}

The phase $\phi (\alpha (t))_{j}$ is given by
\begin{equation*}
\phi (\alpha (t))_{j}=-\frac{i}{\hbar
}\int_{0}^{t}\mathcal{L}[\alpha ,\alpha ^{\ast }]dt^{\prime }
\end{equation*}
where $\mathcal{L}[\alpha ,\alpha ^{\ast }]$ is the classical
Lagrangian of the (independent) systems. In equation (\ref{USC}) we
chose $\omega _{j}(t)=0 $\footnote{$\omega_{j}$ refers to j-th
system.}, since this can be done without violating any of the rules
imposed to construct $\widehat{H}_{sc}(\alpha (t))$. It is just a
matter of simplicity and can also be included, see
\cite{Adelcio2003}. In this case we are left with

\bigskip

\begin{equation}
\widehat{H}(\alpha (t))=\frac{d\alpha ^{\ast }(t)}{dt}(\widehat{a}-\alpha
(t))-\frac{d\alpha (t)}{dt}(\widehat{a}^{\dagger }-\alpha ^{\ast }(t)).
\label{HSC}
\end{equation}
$\alpha $ stands for the $\alpha _{i}$ of the degrees of freedom in question.

The generalization for SU(2) algebra or for any subspace where coherent
states can be included, is immediate. The action of the semiclassical
evolution operator on a coherent state can always be written as\cite{Gilmore}

\begin{equation}
\hat{U}_{sc}^{j}(\alpha (t))\left| \beta _{j}\right\rangle =e^{i\rho
_{j} (t)}\left| \alpha _{j}(t)+\beta _{j}-\alpha
_{j}(0)\right\rangle \label{D  est coe quanquer}
\end{equation}

where $\rho _{j} (\alpha _{j}(t),\beta _{j})=\phi _{j} (\alpha (t))-\mathop{\rm Im}%
\nolimits(\alpha _{j}(t)\beta _{j}^{\ast })+\mathop{\rm
Im}\nolimits\left[ (\alpha _{j}(0)-\beta _{j})^{\ast }\alpha
(t)\right] .$ In general case we have  $\hat{U}_{sc}^{j}(\alpha
(t))=\hat{U}_{sc}^{j}(\alpha_{1} (t),\alpha_{2} (t),\alpha_{3}
(t),...,\alpha_{N} (t))$.

\section{\protect\bigskip\ Time Evolution}

Lets consider a two degrees of freedom system, whose complete
Hamiltonian reads

\begin{equation}
\widehat{H}=\widehat{H}_{1}+\widehat{H}_{2}+\widehat{H}_{12}
\label{hamiltonian}
\end{equation}
where $\widehat{H}_{1(2)}$ represent the autonomous dynamics of the degree
of freedom 1 (2) and $\widehat{H}_{12}$ is their interaction. The
semiclassical Hamiltonian has the following form $\widehat{H}_{sc}(\alpha
,\beta ,t)=\widehat{H}_{sc}^{1}(\alpha ,\beta ,t)+\widehat{H}%
_{sc}^{2}(\alpha ,\beta ,t),$ \footnote {$\widehat{H}_{sc}^{1}(\alpha ,\beta ,t)$ represents the semiclassical Hamiltonian in the subspace 1, and $\widehat{H}_{sc}^{}(\alpha ,\beta ,t)$ represents the semiclassical Hamiltonian in the subspace 2.}and by definition we have $\left[ \widehat{H}%
_{sc}^{1}(\alpha ,\beta ,t),\widehat{H}_{sc}^{2}(\alpha ,\beta ,t)\right]
=0. $ As discussed in section II we rewrite the Hamiltonian (\ref{hamiltonian}) in the following form $\widehat{H}=\widehat{H}_{sc}(\alpha
,\beta ,t)+\widehat{\delta }(\alpha ,\beta ,t),$ where $\widehat{\delta }%
(\alpha ,\beta ,t)=\widehat{H}-\widehat{H}_{sc}(\alpha ,\beta ,t)$ will be
considered as a small perturbation. Using Schr\"{o}dinger's equation we have

\begin{equation*}
i\hbar \frac{\partial }{\partial t}\left| \Psi \right\rangle =\widehat{H}%
_{sc}(\alpha ,\beta ,t)\left| \Psi \right\rangle +\widehat{\delta }(\alpha
,\beta ,t)\left| \Psi \right\rangle
\end{equation*}

where we will always use $\left| \Psi (0)\right\rangle =\left| \alpha
(0)\right\rangle \left| \beta (0)\right\rangle ,$ $\left| \alpha
(0)\right\rangle $ and $\left| \beta (0)\right\rangle $ are coherent states.
It is important to note that by construction we have \footnote{ Since we have builded the semiclassical hamiltonian in such way that $\left\langle\widehat{H}\right\rangle=\left\langle\widehat{H}_{sc}\right\rangle$.}
\begin{equation*}
\left\langle \beta (t)\right| \left\langle \alpha (t)\right| \widehat{\delta
}(\alpha ,\beta ,t)\left| \alpha (t)\right\rangle \left| \beta
(t)\right\rangle =0.
\end{equation*}
Thus, after some straightforward\cite{Adelcio2003} algebraic manipulations we get
\bigskip
\begin{equation}
\left| \Psi \right\rangle (t)=\hat{U}_{sc}(\alpha ,\beta ,t)\left\{
\begin{array}{c}
1+\frac{1}{i\hbar }\int_{0}^{t}\widehat{\Delta }_{s}(t)dt_{1} \\
+\left( \frac{1}{i\hbar }\right) ^{2}\int_{0}^{t}\int_{0}^{t_{1}}\widehat{%
\Delta }_{s}(t_{1})\widehat{\Delta }_{s}(t_{2})dt_{1}dt_{2}+\cdot \cdot \cdot%
\end{array}
\right\} \left| \Psi \right\rangle (0)  \label{est ket geral}
\end{equation}

where $\widehat{\Delta }_{s}(t)$ $=\hat{U}_{sc}^{\dagger }(\alpha ,\beta ,t)%
\widehat{\delta }(\alpha ,\beta ,t)\hat{U}_{sc}(\alpha ,\beta ,t).$

\section{\protect\bigskip Scalar product of States}

Once working in a Hilbert space with a Hermitian Hamiltonian it is
known that the scalar product of any two states which evolved with
the same Hamiltonian must remain constant. This \textquotedblleft
constancy\textquotedblright\ can be a test of our semiclassical
approximation; moreover it  teaches us something \ about \emph{when}
quantum corrections become dominant. The question we address is the
following : let us consider two neighboring states in the sense that
$\left\langle \Psi \mid \Phi \right\rangle \approx 1,$ we know that
$\left\langle \Psi \mid
 \Phi \right\rangle (t)=\left\langle \Psi \mid \Phi \right\rangle (0)$. The important issue is: which of the ingredients of
the quantum evolution are the ones that most affect this relation? Writing
the states $\left\vert \Psi \right\rangle $ and $\left\vert \Phi
\right\rangle $ in the form (\ref{est ket geral}) it is a simple matter to
obtain a semiclassical expansion for $\left\langle \Psi (t)\mid \Phi (t)\right\rangle ,$ for initial states $\left\vert \Psi
(0)\right\rangle =\left\vert \alpha _{0},\beta _{0}\right\rangle $ and $%
\left\vert \Phi (0)\right\rangle =\left\vert \alpha _{0}^{\prime },\beta
_{0}^{\prime }\right\rangle $. For short times we get

\begin{eqnarray}
\left\langle \Psi (t)\mid \Phi (t)\right\rangle =\left\langle
\beta _{0},\alpha _{0}\right| \hat{U}_{sc}^{\dagger }\hat{U}_{sc}^{\prime
}\left| \alpha _{0}^{\prime },\beta _{0}^{\prime }\right\rangle -\frac{1}{%
i\hbar }\int_{0}^{t}dt_{1}\left\langle \beta _{0},\alpha _{0}\right|
\widehat{\Delta }_{1}\hat{U}_{sc}^{\dagger }\hat{U}_{sc}^{\prime }\left|
\alpha _{0}^{\prime },\beta _{0}^{\prime }\right\rangle &&  \notag \\
+\frac{1}{i\hbar }\left( \int_{0}^{t}dt_{1}\left\langle \beta _{0},\alpha
_{0}\right| \hat{U}_{sc}^{\dagger }\hat{U}_{sc}^{\prime }\widehat{\Delta }%
_{1}\left| \alpha _{0}^{\prime },\beta _{0}^{\prime }\right\rangle \right)
&&+\cdot \cdot \cdot  \label{prod1}
\end{eqnarray}

\bigskip where $\hat{U}_{sc}^{\dagger }\hat{U}_{sc}^{\prime }=\hat{U}%
_{sc}^{\dagger }(t,\alpha ,\beta )\hat{U}_{sc}(t,\alpha ^{\prime },\beta
^{\prime })$, e $\widehat{\Delta }_{1}=\widehat{\Delta }_{s}(t_{1,}\alpha
,\beta ).$ Now introducing the resolution of unity in terms of coherent
states $I=I_{1}\otimes I_{2}=\int \frac{d^{2}\zeta _{i}}{\pi }\int \frac{%
d^{2}\xi _{i}}{\pi }\left| \xi _{i},\zeta _{i}\right\rangle \left\langle
\zeta _{i},\xi _{i}\right| $ we get

\begin{eqnarray*}
\left\langle \Psi (t)\mid  \Phi (t)\right\rangle =\left\langle
\beta _{0},\alpha _{0}\right| \hat{U}_{sc}^{\dagger } &&\hat{U}_{sc}^{\prime
}\left| \alpha _{0}^{\prime },\beta _{0}^{\prime }\right\rangle \\
&&-\frac{1}{i\hbar }\int \frac{d^{2}\xi _{i}}{\pi }\int \frac{d^{2}\zeta _{i}%
}{\pi }\left\langle \zeta _{i},\xi _{i}\right| \hat{U}_{sc}^{\dagger }\hat{U}%
_{sc}^{\prime }\left| \alpha _{0}^{\prime },\beta _{0}^{\prime
}\right\rangle \int_{0}^{t}dt_{1}\left\langle \beta _{0},\alpha _{0}\right|
\widehat{\Delta }_{1}\left| \xi _{i},\zeta _{i}\right\rangle \\
+ &&\frac{1}{i\hbar }\int \frac{d^{2}\zeta _{i}}{\pi }\int \frac{d^{2}\xi
_{i}}{\pi }\left\langle \beta _{0},\alpha _{0}\right| \hat{U}_{sc}^{\dagger }%
\hat{U}_{sc}^{\prime }\left| \xi _{i},\zeta _{i}\right\rangle
\int_{0}^{t}dt_{1}\left\langle \zeta _{i},\xi _{i}\right| \widehat{\Delta }%
_{1}^{\prime }\left| \alpha _{0}^{\prime },\beta _{0}^{\prime }\right\rangle
+\cdot \cdot \cdot
\end{eqnarray*}

 The first term reads

\begin{eqnarray}
\left\langle \beta _{0},\alpha _{0}\right| \hat{U}_{sc}^{\dagger }\hat{U}%
_{sc}^{\prime }\left| \alpha _{0}^{\prime },\beta _{0}^{\prime
}\right\rangle &=&\exp \left[ -\frac{\left( \left| \alpha _{t}-\alpha
_{t}{}^{\prime }\right| ^{2}\right) }{2}+i\mathop{\rm Im}\nolimits\left(
\alpha _{t}\alpha _{t}^{\prime \ast }\right) +i(\phi (\alpha ^{\prime
}(t)-\phi (\alpha (t)))\right]  \notag \\
&&\times \exp \left[ -\frac{\left( \left| \beta _{t}-\beta _{t}^{\prime
}\right| ^{2}\right) }{2}+i\mathop{\rm Im}\nolimits\left( \beta _{t}\beta
_{t}^{\prime \ast }\right) +i(\phi (\beta ^{\prime }(t)-\phi (\beta (t)))%
\right] .  \notag
\end{eqnarray}

We note immediately that this equation is relevant for the expansion
\ for t where \ the conditions $\alpha _{t}\approx \alpha
_{t}^{\prime }$ and $\beta _{t}\approx \beta _{t}^{\prime }$ hold.
This t will be longer (almost always) in the case when the system is
regular.

Lets take a look on the time evolution of an operator $a$. We can
compare it with his classical analogue,$a _{cl}$. Suppose we can
write the quantum expectation value as a function of his classical
analogue, such as
\begin{equation*}
\left\langle a\right\rangle _{q}(t)=a _{cl}(t) \left[ 1+F_{q}\left(
t\right) \right]
\end{equation*}
here, $F_{q}\left( t\right) $ is the quantum correction.  Writing $%
F_{q}\left( t\right) $\footnote{$F_{q}(t)$ This function in general
depends on the initial state but is common to have a null first
order. } as
\begin{equation*}
F_{q}\left( t\right) =\sum_{m=2}^{\infty }\frac{1}{m!}\frac{\partial
^{m}F_{q}\left( t\right) }{\partial t^{m}}t^{m}
\end{equation*}
$F_{q}(t)$ then one defines the Erhenfest's time Time as
\begin{equation}
T_{E}=\left( \frac{1}{2}\frac{\partial ^{2}F_{q}\left( t\right)
}{\partial t^{2}}\right) ^{-\frac{1}{2}}.  \label{tempo de
separação}
\end{equation}

The Erhenfest's time as it was described above, means the time where
the quantum corrections are huge enough to become of the same order
of the classical value.

\begin{figure}[th]
\hspace{-3cm}
\vspace{-1cm}
\includegraphics[scale=0.6,angle=270] {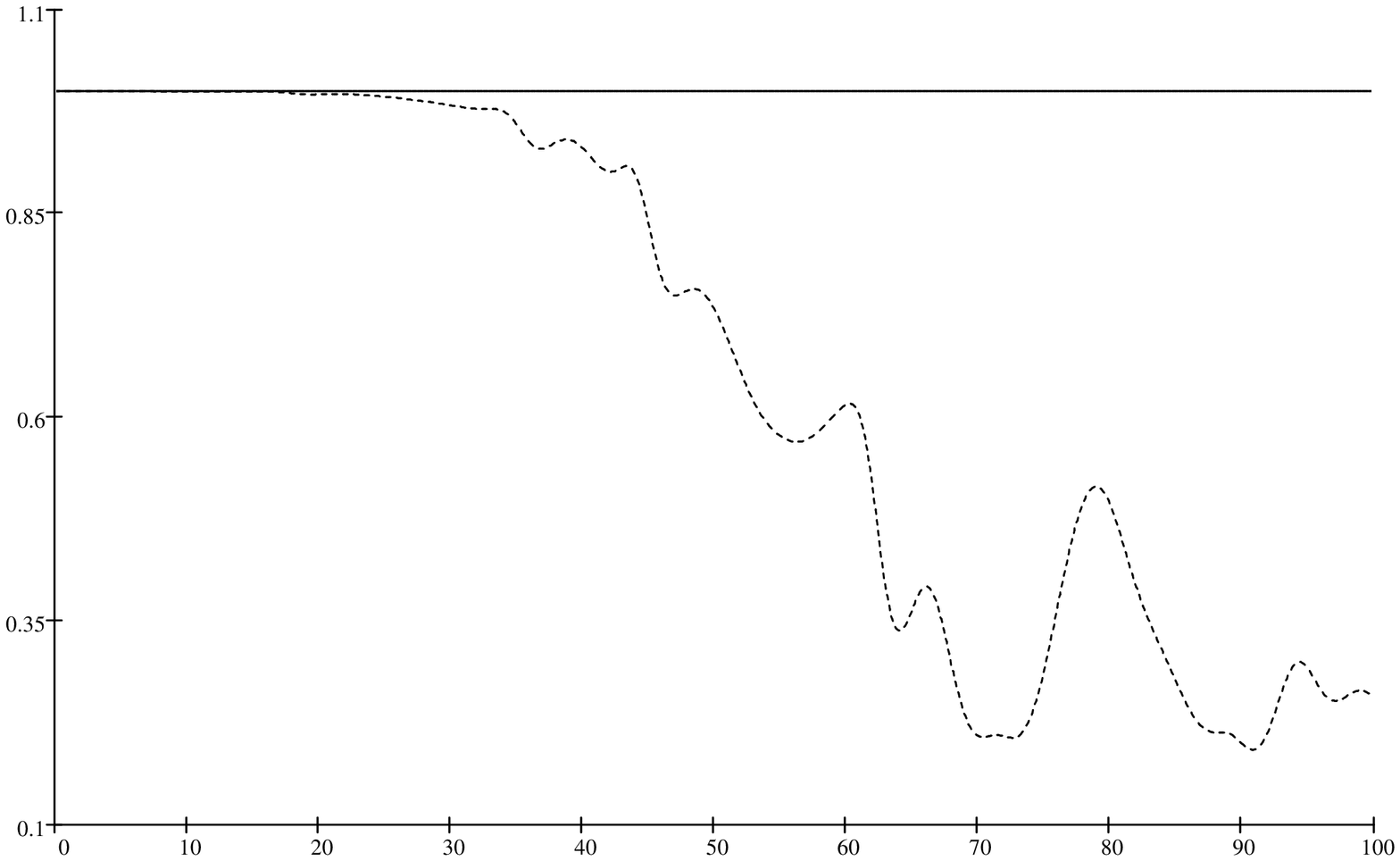}
%\centering
\caption{ Squared modulus of the overlap between two neighbouring states
$\left|\left\langle \alpha _{0}\right| \hat{U}_{sc}^{\dagger }\hat{U}%
_{sc}^{\prime }\left| \alpha _{0}^{\prime }\right\rangle \right|^{2}$, for the driven oscillator, $\ddot{x}+x^{3}=\beta sin(\omega t)$, with $\beta =1$. Chaotic initial conditions (continuous line), $\omega=1.88$, $(x_{1},\dot{x}_{1})=(0,0), (x_{2},\dot{x}_{2})=(0.002,0)$. Regular, same initial conditions (dotted line) for $\omega=3.88$. X axis corresponds to time.}

\label{fig1}
\end{figure}

If we define a semiclassical time $(\tau _{sc})$ as the time during
which we have $\left\langle \Psi (t)\mid  \Phi (t)\right\rangle
\approx
\left\langle \beta _{0},\alpha _{0}\right| \hat{U}_{sc}^{\dagger }\hat{U}%
_{sc}^{\prime }\left| \alpha _{0}^{\prime },\beta _{0}^{\prime
}\right\rangle ,$ then by definition \cite{Adelcio2003} we have
$T_{E}\propto \tau _{sc}$, where $T_{E}$ being Erhenfest's time. Let
us next consider systems with mixed dynamics, e. g. \cite{Karen}. In
this case we know that for the regular regions, neighboring
trajectories keep close for a longer time than trajectories located
in the ergodic region. The
product $\left\langle \beta _{0},\alpha _{0}\right| \hat{U}_{sc}^{\dagger }%
\hat{U}_{sc}^{\prime }\left| \alpha _{0}^{\prime },\beta _{0}^{\prime
}\right\rangle $ will tend more rapidly to zero when we are in a region where there is
chaos \cite{Karen}. Figure (\ref{fig1}) shows the semiclassical square modulus
of the overlap between two neighbouring states, $\left\langle \alpha _{0}\right| \hat{U}_{sc}^{\dagger }\hat{U}%
_{sc}^{\prime }\left| \alpha _{0}^{\prime }\right\rangle ,$ for the
driven conservative oscillator \cite{Gottlieb}. As we can see in
this figure, the behavior of the semiclassical overlap is strongly
dependent of the classical regime, as expected. Thus we can say that
the validity of the semiclassical approximation is longer in the
classical regular conditions.

Consider two classical different initial conditions. Let  $D(t)$ be
the distance of these trajectories in phase space. By definition,
for t=T, $D(T)>>DM \Longrightarrow  e^{-|D(T)|^2}<<1$,  where
$DM=D(\tau_{sc})$, i.e. it is the maximum distance in phase space
that allows us to consider the first term as the most relevant. It
means that there is an appreciable value for the product.
 Under these considerations, we can say that in the chaotic regime we have \footnote{See Appendix.} $\tau
_{sc}\propto \frac{1}{\lambda }\ln (\frac{DM}{D(0)}),$ where
$\lambda$ \footnote{ This exponent is calculated for short time
series \cite{Zeng}.} corresponds to the largest  short time Lyapunov
exponent. If we write $D_{0}$ in terms of the canonically conjugate
variables $q,p$ then

\begin{equation}
D_{0}^{2}=\left| \alpha _{0}-\alpha _{0}{}^{\prime }\right|^{2} =\sqrt{\frac{1}{%
\hbar \omega }}\left[ \frac{m\omega ^{2}}{2}\left(
q_{0}-q_{0}^{\prime }\right) ^{2}+\frac{\left( p_{0}-p_{0}^{\prime
}\right) ^{2}}{2m}\right] .  \label{D 0}
\end{equation}

We can thus say that $D_{0}^{2}\propto S_{cl}$, where $S_{cl}$\ is
the classical action of the system. This result is analogous to
previously obtained ones \cite{Ber77,Ber81}. For regular regions we
have a power law separation of neighboring trajectories,
$D(t)\approx D(0)t^{n}$, thus $\tau _{sc}\propto \left[
\frac{DM}{D(0)}\right] ^{1/n}. $

 We can thus conclude that since the classical trajectories remain
closer for longer times in the case of integrable systems, then we
can say that they are more ``robust" regarding quantum correlations,
i.e., the following terms in the expansion. Chaotic systems on the
contrary will very soon need quantum corrections for an adequate
description.

\bigskip

\section{\protect\bigskip Correspondence Principle Aspects}

The phenomenon baptized as ``scars'' refers to hallmarks of the underlying
classical theory on its quantum counterpart, such as classical periodic
orbits being very conspicuous in Husimi distributions. It was demonstrated
that the existence of scars can be shown by using semiclassical methods \cite%
{Heller84,Heller93}, although the validity of these methods in the chaotic
regime is not known, e.g. we know that WKB fails near caustic points\cite%
{Berry1978}. In this section we will make use of the semiclassical expansion
previously defined and show how it can shed light on the issue.

\subsection{ Husimi's Quantum Phase Space Distribution}

The Q-function or Husimi's function is, see ref.
\cite{Gutz,Schleich} , defined by:
\begin{equation}
\textbf{H}_{\Psi }(Q,P)=\left\langle \alpha \right| \rho _{\Psi }\left| \alpha
\right\rangle ,  \label{eq-1b}
\end{equation}

$\rho _{\Psi }$ is a density operator, and $\left| \alpha \right\rangle $ is
the harmonic coherent state according to the definitions:

\begin{eqnarray}
Q &=&\frac{\left\langle q\right\rangle }{\beta },\text{ \ }P=\frac{%
\left\langle p\right\rangle }{\hbar }\beta  \label{eq-2b} \\
\beta &=&\sqrt{\frac{\hbar }{m\omega }},\alpha =\frac{Q+iP}{\sqrt{2}}.
\label{eq-3b}
\end{eqnarray}

\bigskip From this definition, we are able to write the Husimi expression as:

\begin{equation}
\textbf{H}_{\Phi }(Q,P)=\left| \iiint \Phi (x,y,z)\chi (x,y,z)dxdydz\right| ^{2}.
\label{eq-4b}
\end{equation}

\bigskip

$\Phi (x,y,z)$ is the studied system eigenfunction, and $\chi (x,y,z)$ is
the harmonic coherent state in three dimension.  $\chi (x,y,z)$ can be written as:\

\begin{equation}
\chi _{\alpha }(x,y,z)=\prod_{i=1}^{3}\mu _{\theta _{i}}\exp \left[ -(\frac{%
x_{i}-\left\langle x_{i}\right\rangle _{\alpha _{i}}}{2\Delta x}%
)^{2}+i\left\langle p_{x_{i}}\right\rangle _{\alpha _{i}}\frac{x_{i}}{\hbar }%
\right] ,  \label{eq-5b}
\end{equation}
\ \

with

\bigskip
\begin{eqnarray}
\bigskip \mu _{\theta _{i}} &=&e^{\frac{^{\alpha _{i}^{\ast 2}-\alpha
_{i}^{2}}}{4}}(\frac{\mu \varpi }{\pi \hbar })^{\frac{1}{4}},  \label{Eq-6b}
\\
\Delta x &=&\Delta y=\Delta z=\sqrt{\frac{\hbar }{2\mu \varpi }}.
\label{eq-6bb}
\end{eqnarray}

For the simplest case of  the Harmonic Oscillator,  using equation
[\ref{eq-1b}], the  Husimi the Function for an eigenstate, n, can be
written as:

\begin{equation}
\textbf{H}_{n}(\alpha)=\frac{e^{-\left|\alpha\right|^{2}}\left|\alpha\right|^{2n}}{(2 \pi \hbar) n!}
\end{equation}

In terms of $Q$ and $P$, we have

\begin{equation}
\textbf{H}_{n }(Q,P)=\frac{e^{-(Q^{2}+P^{2})/2}(Q^{2}+P^{2})^{n}}{(4 \pi \hbar) n!}
\end{equation}

\subsection {Husimi Function for the Morse Potential}
The Morse potential is defined as :

\begin{equation}
U(x)=D(e^{-2\alpha x}-2e^{-\alpha x}),  \label{eq-1}
\end{equation}
\ \ where we have defined

\bigskip\ \
\begin{equation}
x\equiv \frac{r-r_{0}}{r_{0}}.  \label{eq-2}
\end{equation}
\ \ \ \ \ \ \ \ \ \ \ \ \ \

\bigskip \qquad The $r_{0}$ values are the equilibrium position of the
center of mass.

$\mu $ is the reduced mass of the two atoms.

$\bigskip $

\bigskip The Hamiltonian that describes the center of mass can be written as:

\bigskip

\begin{equation}
H=\frac{1}{2}\mu (\frac{dr}{dt})^{2}+\frac{1}{2}\frac{L^{2}}{\mu r^{2}}+U(r).
\label{eq-3}
\end{equation}
\ \ \
The time independent  Schr\"{o}dinger equation is:\ \ \ $%
\bigskip $%
\begin{equation}
\lbrack -\frac{\hbar ^{2}}{2\mu }\frac{1}{r}\frac{\partial ^{2}}{\partial
r^{2}}r+\frac{L^{2}}{2\mu r^{2}}+V(r)]\Phi (r,\theta ,\varphi )=E\Phi
(r,\theta ,\varphi ).  \label{eq-13}
\end{equation}
\ \

We can write the wavefunction as

\begin{equation}
\Phi (r,\theta ,\varphi )=\frac{\Psi (r)}{r}Y_{lm}(\theta ,\varphi ),\
\label{eq-14}
\end{equation}
\ \ \

\bigskip where $Y_{lm}$ is the spherical harmonics, so that is:

\ \
\begin{equation}
Y_{l}^{m}(\theta ,\varphi )=A_{lm}P_{l}^{m}[\cos (\theta )]e^{im\varphi }.
\label{eq-15}
\end{equation}
\

For L=0 case we find the eingenvalues:
\begin{equation}
E=-D+\hbar \varpi \lbrack (\nu +\frac{1}{2})-\frac{1}{\zeta }(\nu +\frac{1}{2%
})^{2}]  \label{eq-16}
\end{equation}

\bigskip and for the eigenfunctions:
\begin{equation}
\frac{\Psi (x)}{r}=\frac{A_{1}}{r_{0}(x+1)}\frac{\Gamma (\varsigma )}{\Gamma
(\tilde{a})}\exp (-\beta _{1}x-\frac{1}{2}\zeta _{2}e^{-\alpha
x})\sum_{n=0}^{\infty }[\frac{\Gamma (\tilde{a}+n)}{\Gamma (\varsigma +n)}%
\frac{(\zeta _{2}e^{-\alpha x})^{n}}{n!}].  \label{eq-17}
\end{equation}

Where \ \ $\nu =0,1,2,3,...$ \ \ \ \ \ $\nu <\frac{1}{2}(\zeta _{2}-1),$ and

\bigskip

\begin{eqnarray}
\beta ^{2} &=&-\frac{2\mu Er_{0}^{2}}{\hbar ^{2}},\text{ \ }\zeta =\frac{%
2\gamma }{\alpha },  \label{eq-18} \\
\zeta _{2} &=&\frac{2\gamma}{\alpha }.,\text{ \ \ }\varsigma =\frac{%
2\beta }{\alpha }+1 \\
\tilde{a} &=&\frac{1}{2}\varsigma -\frac{\gamma }{\alpha}%
\ .
\end{eqnarray}

\ \ \ \ \ \ \ \ A$_{1}$ is fixed by normalization.

Following the definition [\ref{Eq-6b}], we obtain the Husimi the Function as
\begin{equation}
\textbf{H}_{\Phi }(Q,P)=\left| 2\pi \int_{-1}^{\infty }I_{\theta 0}(x)\varkappa
(x)\Psi (x)r_{0}^{2}(x+1)dx\right| ^{2}.
\end{equation}
\ \ \ \ \
where $I_{\theta 0}(r) =\frac{1}{\sqrt{\pi }}\frac{\sinh (\alpha _{0}r)}{\alpha
_{0}r},\text{ \ }l=0$.

\subsection{ Semiclassical Husimi's Function }

The semiclassical expansion, as defined above, give us a time evolution of a
quantum state as a pertubative expansion. A eigenstate has only a time
dependent phase as its dynamics. The nearest semiclassical scenario we can
build is to choose a coherent state with the same energy. The time
dependence can be eliminated by a time integration, i.e. mean in time. This
integration can be justified noting that as we are dealing with eigenstates
we have not time precision.

Under this considerations\footnote{ In case of classical mixed
dynamics we must perform a mean considering all possible initial
condition for the specific energy.} we may write the semiclassical
Husimi function as

\begin{equation}
\textbf{H}_{sc}(\beta )\approx \overline{\left\vert \left\langle \beta |\alpha
(t)\right\rangle \right\vert }^{2}.  \label{hu}
\end{equation}

The states  $\left|\alpha (t)\right\rangle$ and $\left|\beta\right\rangle$ are coherent states of the harmonic oscillator. $\alpha(t)$ is defined as $\alpha(t)= \frac{Q(t)+iP(t)}{2}$
and $\beta=x+ip_{x}$, where $x$ and $p_{x}$ are parameters of the Husimi Function.

$Q(t)$ and $P(t)$ are the classical canonical conjugate pairs.

Easily we can show  that
\begin{equation}
\textbf{H}_{sc}(\beta )\approx \overline{e^{\left|Q(t)+iP(t)-\beta\right|^{2} }}
\label{hu2}
\end{equation}

For the morse potential, with L=0, we obtain the classical trajectory \cite{Oliveira01}:
\begin{equation}
x(t)=\frac{1}{\alpha }\ln \{\frac{D}{\mid E\mid }+[\sin (t%
\sqrt{\frac{2\mid E\mid \alpha ^{2}}{\mu r_{0}^{2}}})\sqrt{\frac{%
D^{2}+ED}{\mid E\mid ^{2}}}]\}\ \ \ E<0
\label{eq-8b2}
\end{equation}

We also have $p(t)=\mu \frac{dx(t)}{dt}$, and we can choose $p(0)=0$ and using $E=E_{n}$ into (\ref{hu2}) to obtain the semiclassical Husimi function.

\begin{figure}[th]
\includegraphics[scale=1.0] {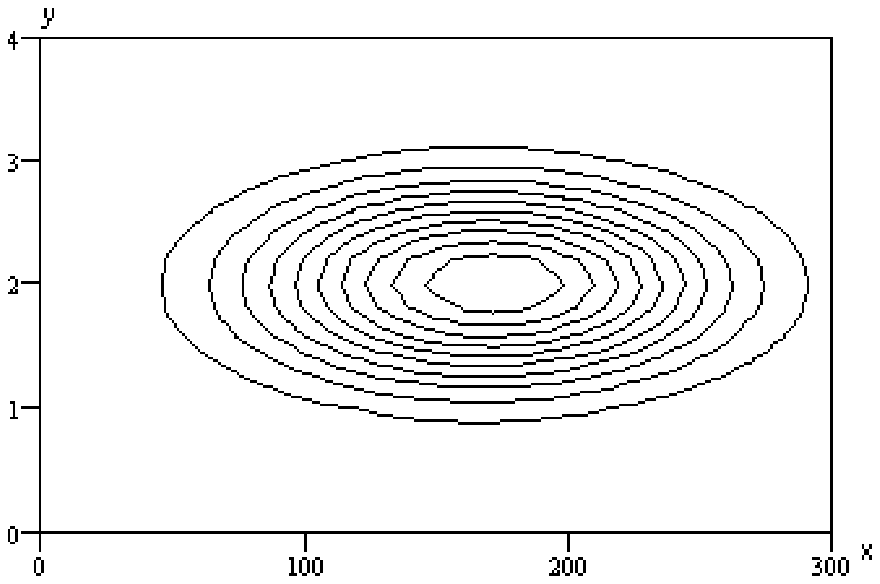} \vspace{2cm}
\caption{ semiclassical Husimi function for the Morse Potential, $U(x)=D(e^{-2%
\protect\alpha x}-2e^{-\protect\alpha x}),$ where $x=\frac{r-r_{0}}{r_{o}}.$
We have used the experimental hydrogen molecule values, $\protect\alpha %
=1.440$ and $D=4.75(eV).$ The principal quantum number n=0 and the angular
momentum L=0. The x axis is related to the position, r, as $r=r_{0}\frac{%
(x+10)}{250}$, the momentum p is related to the y axis variable as $p=m%
\protect\omega r_{0}\frac{(y-2)}{1.2}.$ We have used $r_{0}=7.42\times
10^{-11}m$ and $\protect\omega =$8.3$\times $10$^{14}$rad/s.}
\label{fig2}
\end{figure}

\begin{figure}[th]
\includegraphics[scale=1.0] {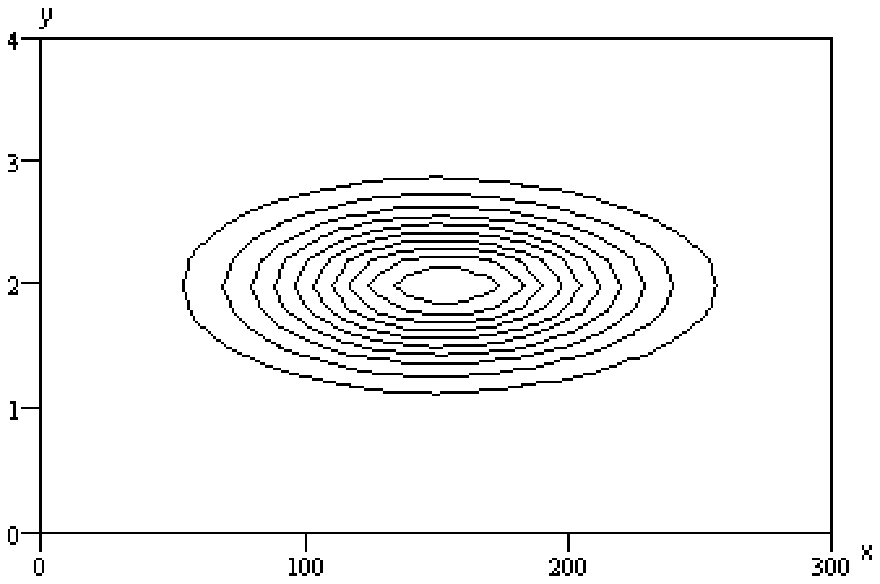} \vspace{2cm}
\caption{Husimi function for the Morse Potential with the same parameters of
figure(\ref{fig2})}
\label{fig3}
\end{figure}

\begin{figure}[ht]
\includegraphics[scale=1] {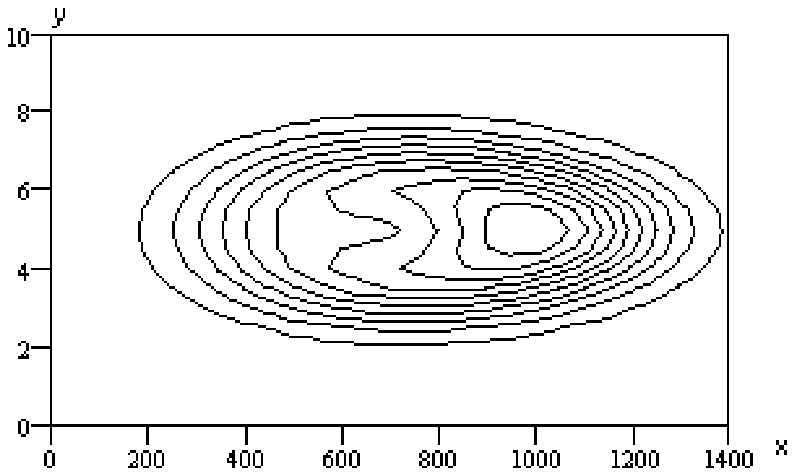} \vspace{1cm}
\caption{Semiclassical Husimi function for the Morse Potential, $U(x)=D(e^{-2%
\protect\alpha x}-2e^{-\protect\alpha x}),$ where $x=\frac{r-r_{0}}{r_{o}}.$
We have used the experimental hydrogen molecule values, $\protect\alpha %
=1.440$ and $D=4.75(eV).$ The principal quantum number n=1 and the angular
momentum L=0. The x axis is related to the position, r, as $r=r_{0}\frac{%
(x+300)}{1000}$, the momentum p is related to the y axis variable as $p=m%
\protect\omega r_{0}\frac{(y-5)}{5}.$ We have used $r_{0}=7.42\times
10^{-11}m$ and $\protect\omega =$8.3$\times $10$^{14}$rad/s }
\label{fig4}
\end{figure}

\begin{figure}[th]
\includegraphics[scale=1] {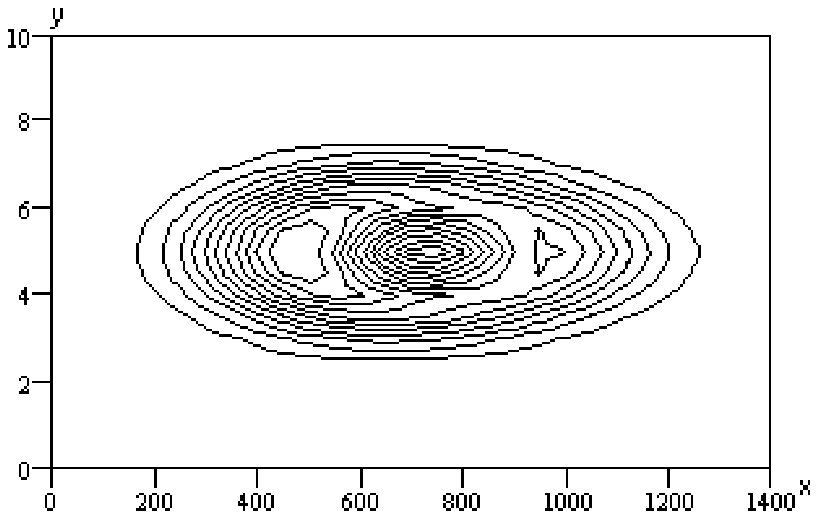} \vspace{1cm}
\caption{Husimi function for the Morse Potential with the same parameters of
figure(\ref{fig4})}
\label{fig5}
\end{figure}

In figure (\ref{fig2}) we show the approximated Husimi for the Morse
potential with the parameters of the $H_{2}$ molecule, for $n=0$. In
figure (\ref{fig3}) we have the exact result, figure (\ref{fig4})
shows the semiclassical Husimi function for n=1 and figure
(\ref{fig5}) the exact result, details about exact calculation can
be found in ref. \cite{Oliveira01} . As we can observe in this
figures, (\ref{fig2}) and (\ref{fig3}), the semiclassical Husimi
function does not reproduce exactly the Husimi function, but it
regards some similarities.

Now consider the Harmonic potential, thus  we have
\begin{equation}
Q(t)=Q(0)cos(\omega t)+\frac{P(0)}{\omega}sin(\omega t)
\label{qt}
\end{equation}

and

\begin{equation}
P(t)=P(0)cos(\omega t)-Q(0)\omega sin(\omega t)
\label{PT}
\end{equation}

We chose a coordinate system such as the Hamiltonian can be written
as
\begin{equation}
H =\omega \frac{(Q^{2}+P^{2})}{2}.
\end{equation}

\begin{figure}[th]
\includegraphics[scale=.35,angle=270] {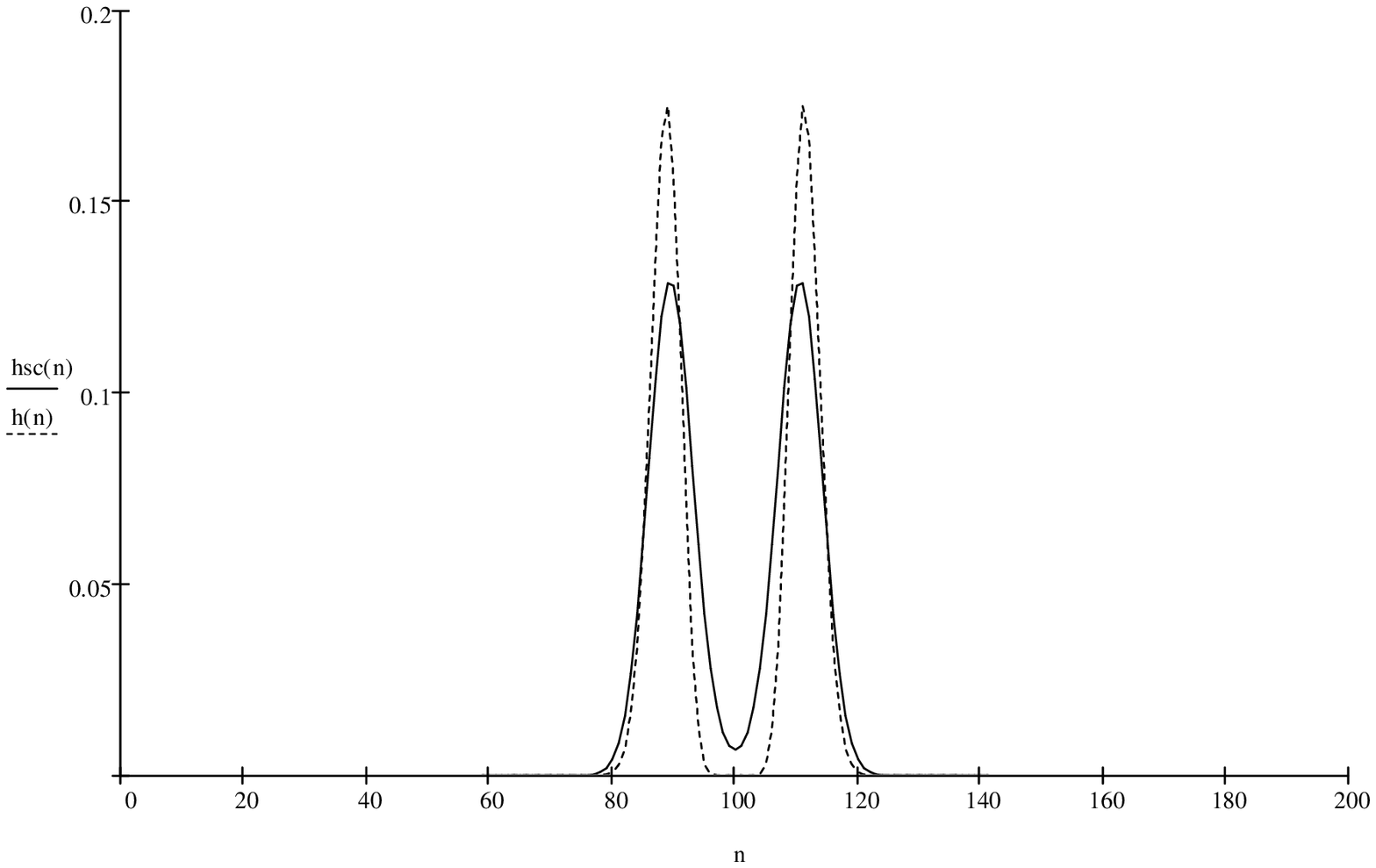} \vspace{0cm}
\caption{Full line shows a cross section of Husimi function for the
harmonic potential for p=0. The principal quantum number n=5. The x
axis is related to the position. Dotted line, correspondent
semiclassical Husimi function. } \label{fig6}
\end{figure}

\begin{figure}[th]
\includegraphics[scale=0.35, angle=270] {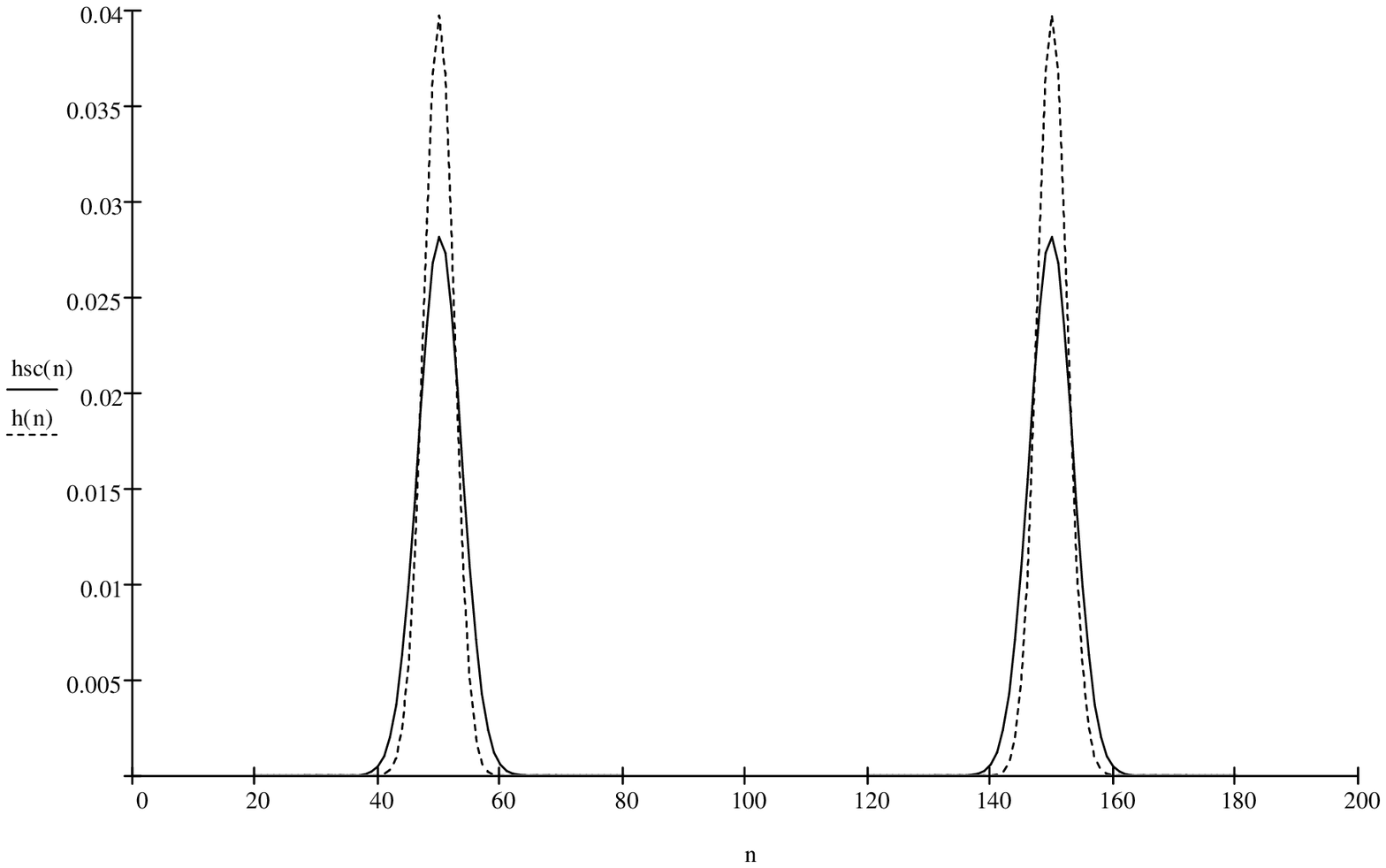} \vspace{0cm}
\caption{Full line shows a cross section of Husimi function for the
harmonic potential for p=0. The principal quantum number n=100. The
x axis is related to the position. Dotted line, correspondent
semiclassical Husimi function.} \label{fig7}
\end{figure}

\begin{figure}[th]
\includegraphics[scale=0.5,angle=270] {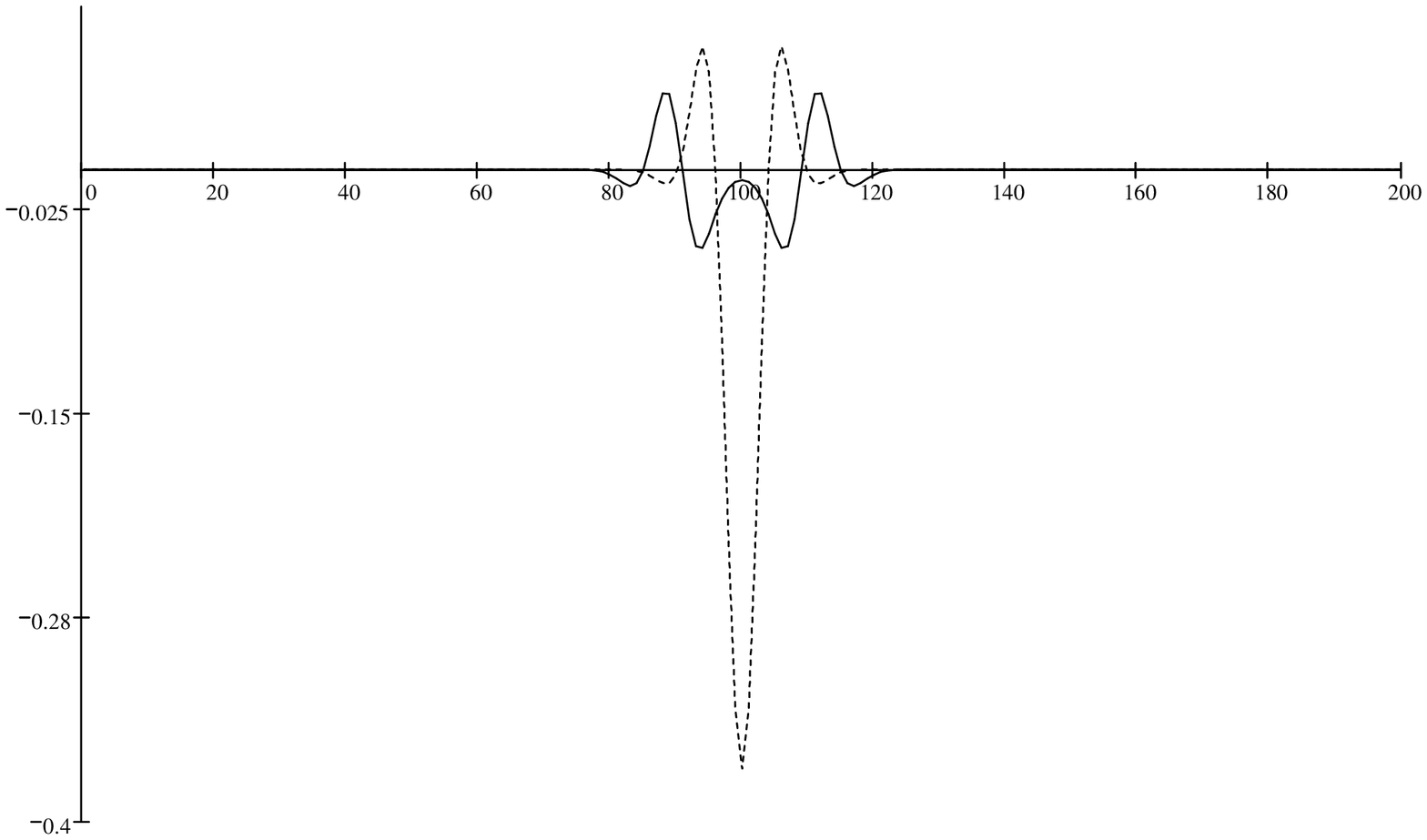} \vspace{-0cm}
\caption{Dotted line shows  $Q_{d}$ function for the harmonic oscillator with the principal quantum number n=1, full line n=5. X axis corresponds to position.}
\label{fig10}
\end{figure}

\begin{figure}[th]
\includegraphics[scale=0.5,angle=270] {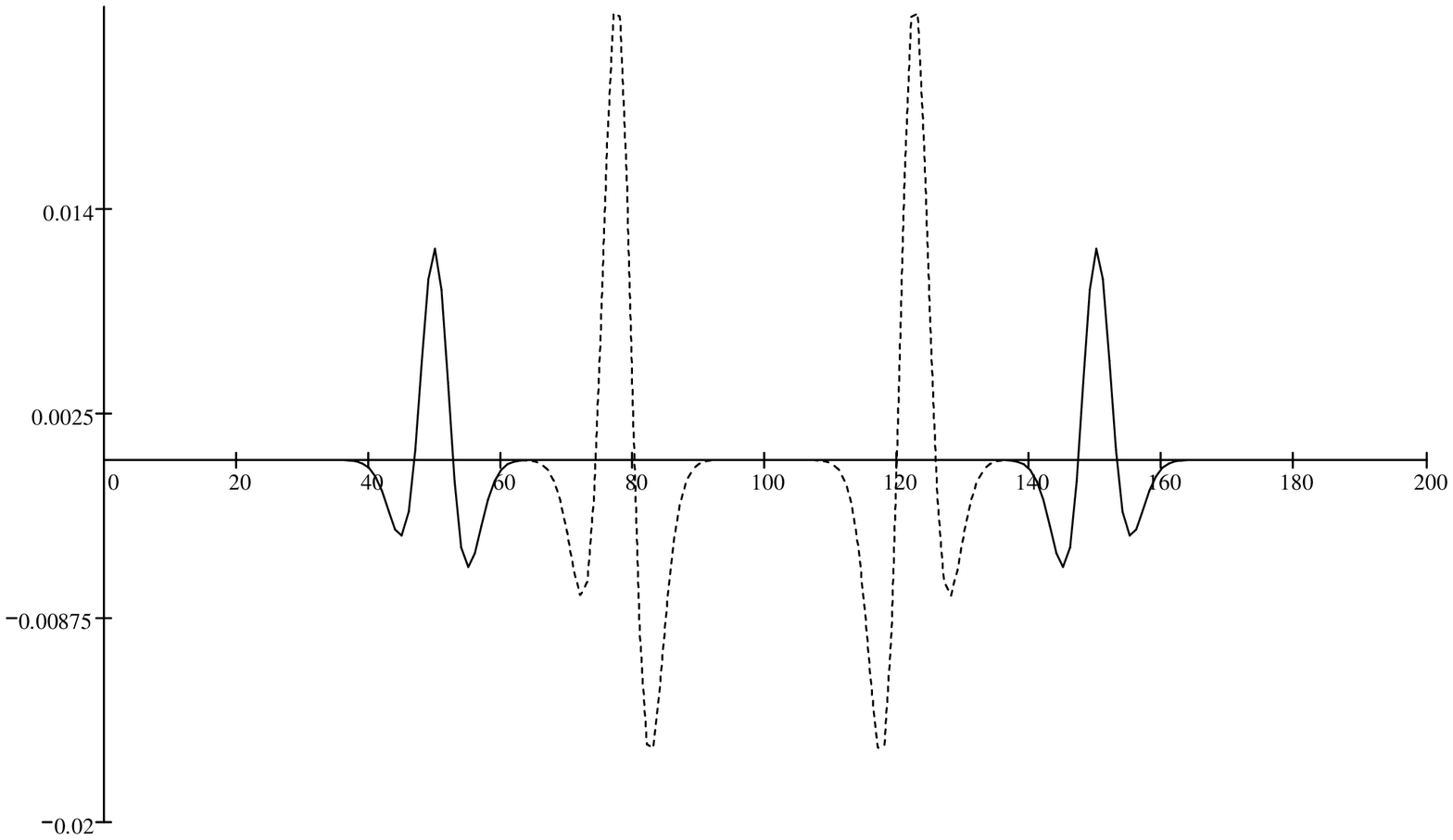} \vspace{-0cm}
\caption{Dotted line shows  $\Delta Q_{d}$ function for the harmonic
oscillator with the principal quantum number n=20, full line n=100.
X axis corresponds to position.} \label{fig11}
\end{figure}

\begin{figure}[th]
\includegraphics[scale=0.5,angle=00] {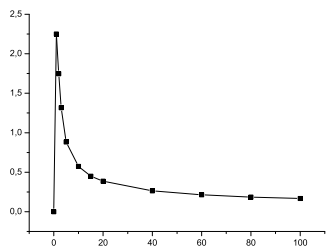} \vspace{-0cm}
\caption{$SQ_{d}$ function for the harmonic oscillator.X axis
corresponds to the principal quantum number n.} \label{fig11b}
\end{figure}

Substituting (\ref{qt}) and  (\ref{PT}) into (\ref{hu2}) we obtain a
semiclassical Husimi Function for the Harmonic oscillator with an
energy $E_{n}=\hbar \omega (n+\frac{1}{2})$, without any lost of
generality we can use  $Q(0)=(2E_{n})^\frac{1}{2}$, and $P(0)=0$. In
figure (\ref{fig6}) we show the approximated and exact Husimi for
the Harmonic potential for n=5. Figure (\ref{fig7}) shows the
semiclassical and exact Husimi function for n=100. In the figures
(\ref{fig3}) to (\ref{fig11}). In order to quantify the quality of
the approximation we define the function $\Delta \textbf{Q}_{\psi
_{n}}(q)$ as
\begin{equation}
\Delta \textbf{Q}_{\psi
_{n}}(q)=\textbf{H(q,p=0)}-\textbf{H(q,p=0)}_{sc},
\end{equation}
where $\textbf{H}$ is the exact Husimi function and
$\textbf{H}_{sc}$ is the semiclassical Husimi function.
 In Figure (\ref{fig10}) we show
$\Delta\textbf{Q}_{\psi _{n}}(q)$, for n=1 and n=5. Figure
(\ref{fig11}) same graph for n=20 and n=100. Due to the spherical
symmetry we have chosen p=0. X axis corresponds to position. As we
increase the principal quantum number (n) we have $\Delta
\textbf{Q}_{\psi _{n}}(q)\rightarrow 0$. In order to see the
classical limit, let us define the function $SQ_{\psi _{n}}$, witch
is

\begin{equation}
SQ_{\psi_{n}}=\int_{-\infty}^{\infty}{\left|{\Delta
Q}_{\psi_{n}}(q)\right|dq} \label{SQ}
\end{equation}
Suppose we have $SQ_{\psi }\approx 0$. It means that quantum
description  of the state $\psi$, in the Husimi's representation, is
almost contained in the semiclassical one. Of course it does not
means that we have no quantum features, it only meas that Husimi is
not a good observable for this situation \cite{Oliveira02}. Although
that we can  say that the quantum classical difference became
smaller, as expected. Figure (\ref{fig11b}) shows $SQ_{n}$ for some
eigenstates of the harmonic oscillator. For n=0 we have a null
$SQ_{0}$, what was expected, since the fundamental harmonic
oscillator eigenstate is a coherent state. Also we can say that the
approximation works better as we increase the principal quantum
number, as expected. From these figures we may conclude that the
classical ingredient is very strong on the state formation of
regular systems.

\section{Semiclassical Fidelity}

 We now make a simple test of our approximation by applying it to a
well known behavior of the fidelity. We know\cite%
{Peres84,Cucchietti01,Benenti2002,Prosen02,Prosen03,Weinstein02} that for a
linear perturbation , fidelity decays, for short times, as \bigskip\ $%
e^{-at^{2}}$ once one starts with Gaussian states. As we will see, this
result is independent of the classical behavior of the system.

Let us consider \ the product
\begin{equation}
f(\Psi ,V)=\left\langle \Psi \right| U_{0}(t)U(t)\left| \Psi \right\rangle
\label{fidelity}
\end{equation}

where $U_{0}=e^{-\frac{i}{\hslash }Ht}$ and $U=e^{-\frac{i}{\hslash }%
(H+V)t}, $ $V$ is a perturbation, and it is given by $V=\varepsilon
\hbar (a+a^{\dagger })$ where we have $\varepsilon \ll 1$. Fidelity
$(F)$ is defined as $F\equiv \left| f(\Psi ,V)\right| ^{2}.$ \ Let
$\alpha (t)$ be the classical evolution of $\alpha (0)$ under the
action of $H_{0\text{ }}$, and
$\beta (t)$ the classical evolution of $\beta (0)$ under the action of $%
H=H_{0}+V.$ We consider $\beta (0)=\alpha (0).$ In this case for short times
we have $\alpha (t)\approx \beta (t)+i\varepsilon t.$ Our semiclassical expansion
gives
\begin{eqnarray*}
f(\alpha ,V) &=&\left\langle \alpha (t)\right| \left| \beta (t)\right\rangle
e^{i\left[ \phi (\beta (t))-\phi (\alpha (t))\right] } \\
&&+\frac{1}{i\hbar }\int_{0}^{t}dt_{1}e^{i\eta _{3}}\left\langle \alpha
(t_{1})\right| \widehat{\delta }[\alpha (t_{1})]\left| \alpha
(t_{1})-i\varepsilon t\right\rangle \\
&&-\frac{1}{i\hbar }\int_{0}^{t}dt_{1}e^{i\eta _{4}}\left\langle \beta
(t_{1})+\varepsilon t\right| \widehat{\delta }[\beta (t_{1})]\left| \beta
(t_{1})\right\rangle +\cdot \cdot \cdot
\end{eqnarray*}

\begin{figure}[th]
\includegraphics[scale=0.34, angle=270] {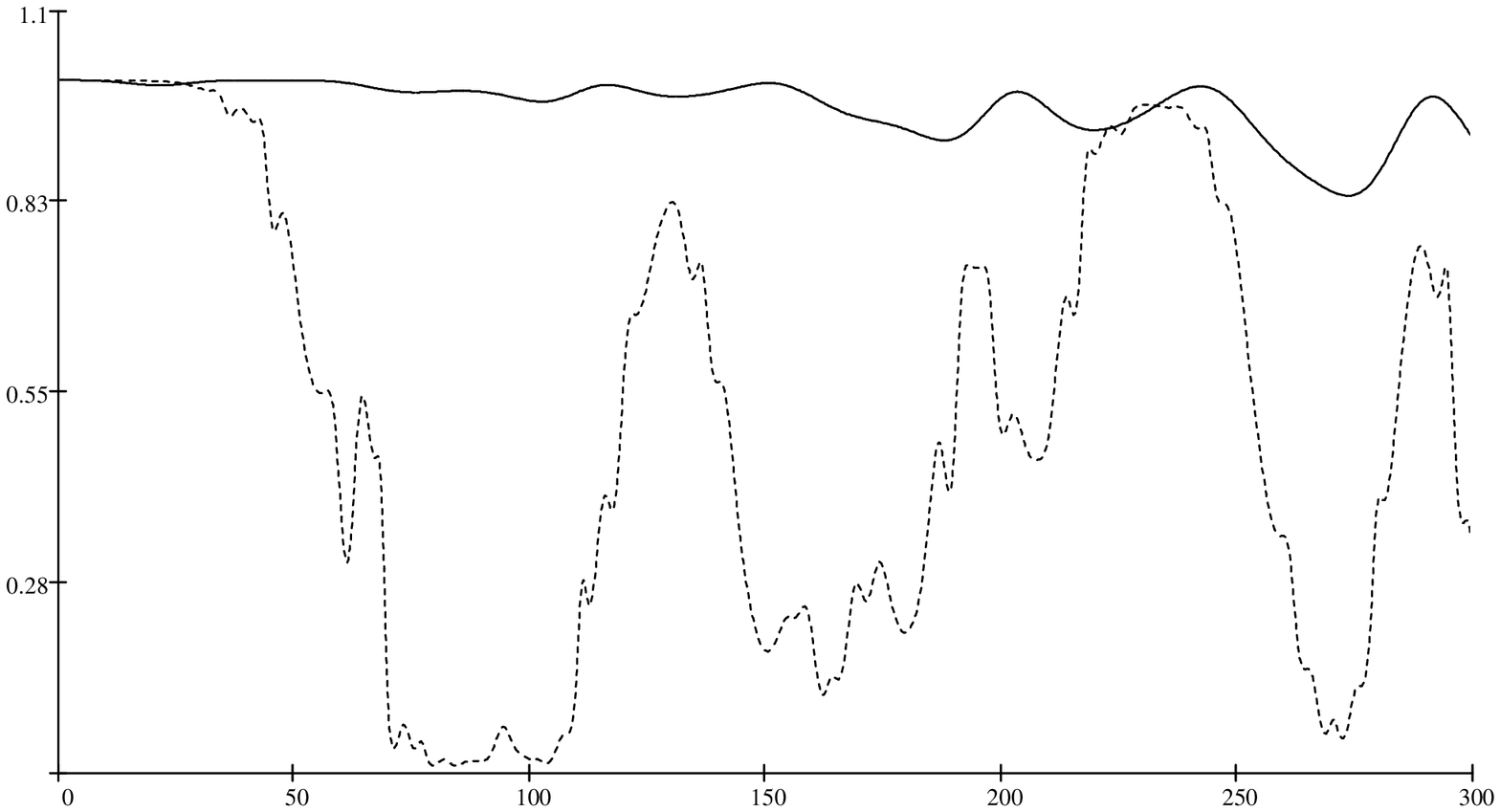} \vspace{-0cm}
\caption{Semiclassical fidelity $\left|\left\langle\alpha(t) \mid
\beta(t)  \right\rangle\right|_{sc}^{2}$, for the driven oscillator,
$\ddot{x}+x^{3}=\gamma sin(\omega t)$, with $\gamma =1$. Chaotic,
$\omega=1.88$, initial condition (dotted line)
$(x_{1},\dot{x}_{1})=(0,0)$. Regular, $\omega=3.88$, same initial
conditions (continuous line). Both curves with the same pertubation
potential. X axis corresponds to time.} \label{fig12}
\end{figure}

\begin{figure}[th]
\includegraphics[scale=0.4,angle=270] {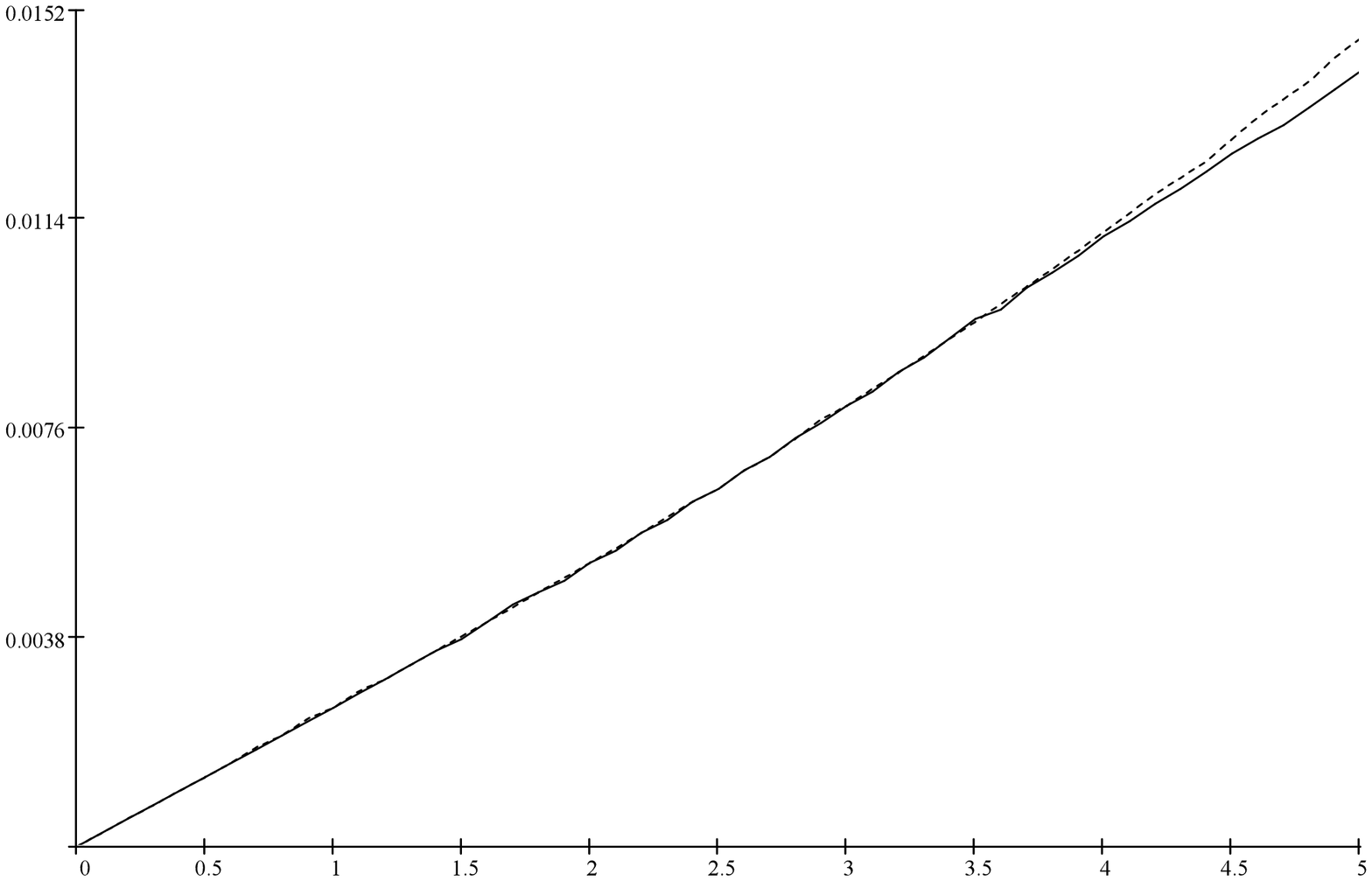} \vspace{-0cm}
\caption{ Semiclassical overlap $\sqrt{ln{\left|
\left\langle\alpha(t) \mid \beta(t) \right\rangle\right|_{sc}^{2}}}$
, for the driven oscillator, $\ddot{x}+x^{3}=\gamma sin(\omega t)$,
with $\gamma =1$. Chaotic, $\omega=1.88$, initial condition (dotted
line) $(x_{1},\dot{x}_{1})=(0,0)$. Regular, $\omega=3.88$, same
initial conditions (continuous line). Both curves with the same
pertubation potential.} \label{fig13}
\end{figure}

The term $\left\langle \alpha (t)\right| \left| \beta (t)\right\rangle $
gives the short time behavior for $F$ and one gets

\begin{equation}
\left\langle \alpha (t)\right\vert \left\vert \beta (t)\right\rangle =\exp
\left[ -\frac{\varepsilon ^{2}t^{2}}{2}-i\varepsilon t\right]  \label{fid 1}
\end{equation}%
The above result is independent of the presence of chaos in the
classical dynamics. In Figure (\ref{fig12}) we show the product
$\left| \left\langle\alpha(t) \mid \beta(t)
\right\rangle\right|_{sc}^{2}$. As we can easely observe, the
chaotic initial condition decay faster. Although that, the short
time dynamics has a gaussian decay, figure (\ref{fig13}) we show
$\sqrt{ln{\left| \left\langle\alpha(t) \mid \beta(t)
\right\rangle\right|_{sc}^{2}}}$ for short time evolution, witch is
a straight line. As the first term in the semiclassical
approximation is valid for longer times for integrable systems than
for chaotic ones, we conclude that the gaussian regime should be
valid for longer times as it was already found
numerically in several examples of the literature\cite%
{Peres84,Prosen02,Prosen03,Weinstein02}.
Let us now look at the general initial state situation, $f(\alpha ,V)$ can
be written as
\begin{equation}
\left\langle \Psi \right| U_{0}(t)U(t)\left| \Psi \right\rangle =\int \frac{%
d^{2}\alpha }{\pi }\int \frac{d^{2}\beta }{\pi }\left\langle \Psi \right|
\left| \alpha \right\rangle \left\langle \alpha \right| U_{0}(t)U(t)\left|
\beta \right\rangle \left\langle \beta \right| \left| \Psi \right\rangle
\label{fid 2}
\end{equation}

In the equation (\ref{fid 2}) we note that when $\alpha =\beta ,$ we have $%
\left\langle \alpha \right| U_{0}(t)U(t)\left| \beta \right\rangle =f(\alpha
,V)$ as defined in (\ref{fid 1}), and it has the same characteristics
discussed above. However, the product $\left\langle \Psi \right| \left|
\alpha \right\rangle \left\langle \beta \right| \left| \Psi \right\rangle
\left\langle \alpha \right| U_{0}(t)U(t)\left| \beta \right\rangle $ is not
expected to be very relevant when $\alpha $ is very different from $\beta ,$
which is certainly true for short times. The whole analysis is weakened by
the fact that the product $\left\langle \beta \mid \Psi
\right\rangle $ may be important in chaotic and regular regions
simultaneously. In this case it becomes very difficult to give a general
estimate for the validity of (\ref{fid 1}).

\section{Conclusions }
As a general remark, we can say that, Overlap, Scar and short time
Fidelity are strongly determined by the semiclassical dynamics,
therefore we can say that the classical imprints are determinant. We
also observe that overlap decay time has a very different dynamics
from fidelity decay, although they are very similar in conception.
In the particular harmonic case, we show that the first
semiclassical term is able to reproduce the Husimi function with a
increasing accuracy as we increase the principal quantum number n.
We must remark that there is no demonstration that would suggest an
existence of a limit procedure witch turns quantum corrections less
important in terms of the proposed semiclassical expansion.

\textbf{Acknowledgments:} The author is grateful to Fapesb for
partial financial  support. The author also acknowledge M. C. Nemes
and Enio Jelihovschi  for helpful comments.

\newpage
\appendix{\textbf{Appendix}}

\section{Semiclassical expansion and the Lyapunov Exponent} The
semiclassical zeroth order is always a coherent state or a tensor
product of coherent states,with the labels are described by
classical dynamics. For a general state we have
\[
\left\langle \psi | \phi \right\rangle =\left\langle \psi | \phi
\right\rangle _{sc}+\text{corrections. }
\]

Consider that initially the states are  $\left| \psi
(0)\right\rangle =\prod_{i}\left| \alpha _{i}(0)\right\rangle $ e
$\left|
\phi (0)\right\rangle =\prod_{i}\left| \beta _{i}(0)\right\rangle ,$ where $%
\left| \alpha _{i}(0)\right\rangle $ e $\left| \beta
_{i}(0)\right\rangle $ are coherent states. Then the zeroth
semiclassical term is $\left\langle \psi (t)| \phi (t)\right\rangle
_{sc}=\prod_{i}\left\langle \alpha _{i}(t)| \beta
_{i}(t)\right\rangle $ . We know \cite{Gilmore}that the overlap of
coherent states is
\begin{eqnarray}
\left| \left\langle \alpha _{i}\right| \left| \beta
_{i}\right\rangle \right| ^{2} &=&\exp (-\left| \alpha _{i}-\beta
_{i}\right| ^{2})\text{ \ \
\ \ \ \ \ \ \ \ \ \ \ \ \ \ \ \ \ \ \ \ \ \ for h(3);}  \label{A1} \\
&=&\left[ 1-\frac{\left| \alpha _{i}-\beta _{i}\right|
^{2}}{(1+\left| \beta _{i}\right| ^{2})(1+\left| \alpha _{i}\right|
^{2})}\right] ^{2J}\text{ \ \ \ for su(2).}  \label{A2}
\end{eqnarray}

The equation (\ref{A2}) coincides with (\ref{A1}) if we make  $%
\alpha _{i}=\alpha _{i}/\sqrt{2J}$ and taking the limit
$J\rightarrow \infty .$ In this situation we can say that
\begin{eqnarray}
-\left| \alpha _{i}-\beta _{i}\right| ^{2} &=&\ln (\left|
\left\langle
\alpha _{i}| \beta _{i}\right\rangle \right| ^{2})  \nonumber \\
\left| \alpha _{i}-\beta _{i}\right| &=&\sqrt{2}(\ln \left|
\left\langle \alpha _{i}| \beta _{i}\right\rangle \right|
^{-1})^{1/2}. \label{A4}
\end{eqnarray}

We can also say that $\left| \alpha _{i}-\beta _{i}\right|
^{2}=(x_{i}-y_{i})^{2}+(p_{xi}-p_{yi})$\bigskip $^{2}\equiv \Delta
q^{2}+\Delta p^{2}$.

The Lyaounov  exponent is defined as

\begin{equation}
\lambda =\lim_{t\rightarrow \infty }\lim_{\Delta x(0)\rightarrow 0}\frac{1}{t%
}\ln \frac{\Delta x(t)}{\Delta x(0)}  \label{A3}
\end{equation}
onde $\Delta x(t)=\left| x_{1}(t)-x_{2}(t)\right| ,$ where
$x_{i}(t)$ is the clasical evolution for $x_{i}(0)$ as initial
condition. In the above limit, we get

\[
\left| \left\langle \alpha _{i}(t)| \beta _{i}(t)\right\rangle
\right| ^{2}=\exp \left[ -\left( \Delta q(0)^{2}e^{2\lambda
_{q}t}+\Delta p(0)^{2}e^{2\lambda _{q}t}\right) \right]
\]
As we have  $\sum_{i}\lambda _{i}=0,$ the biggest Lyapunov
exponent($\lambda _{\max })$ is approximately

\bigskip\
\[
\lambda _{\max }=\lim_{t\rightarrow \infty }\lim_{\beta
_{i}(0)\rightarrow \alpha _{i}(0)}\frac{1}{2t}\ln \left[ \frac{\ln
\left| \prod_{i}\left\langle \alpha _{i}(t) | \beta
_{i}(t)\right\rangle \right| ^{-1}}{\ln \left| \prod_{i}\left\langle
\alpha _{i}(0) | \beta _{i}(0)\right\rangle \right| ^{-1}}\right] .
\]
Observing this equation we may say that the Lyapunov exponent is
related with the quantum nature of the system. As faster the quantum
corrections are needed, i.e. how faster the product  $\left|
\left\langle \alpha _{i}(t)| \beta _{i}(t)\right\rangle \right|
\rightarrow 0$ , bigger is the Lyapunov exponent. This behavior has
already been pointed by many others \cite{Ber81,Ber77}.

\newpage

%\newpage

%\newpage
%\newpage

%\newpage

%\newpage

\newpage

\bigskip

\end{document}